\pdfoutput=1
\documentclass[3p,a4paper,times,12pt]{elsarticle}

\usepackage{graphicx}
\usepackage{amsmath}
\usepackage{makecell}
\usepackage{hyperref}

\newcommand{\ie}{{\it i.\,e.} }

\begin{document}

\begin{frontmatter}
\title{Disturbances in the Doppler frequency shift of ionospheric signal and in telluric current caused by the atmospheric waves from an explosive eruption of Hunga Tonga volcano on January 15, 2022}

\author[ion]{N.\,Salikhov}
\author[lpi,ion]{A.\,Shepetov}
 \ead{ashep@www.tien-shan.org}
\author[ion]{G.\,Pak}
\author[ion,afi]{V.\,Saveliev}
\author[ion]{S.\,Nurakynov}
\author[lpi]{V.\,Ryabov}
\author[lpi]{V.\,Zhukov}

\address[ion]{Institute of Ionosphere, Kamenskoye plato, Almaty, Republic of Kazakhstan, 050020}
\address[lpi]{P.\,N.\,Lebedev Physical Institute of the Russian Academy of Sciences (LPI), Leninsky pr., 53, Moscow, Russia, 119991}
\address[afi]{V.\,G.\,Fesenkov Astrophysical Institute, Almaty, 050020, Kazakhstan}

\date{March 7, 2023}

\begin{abstract}
After an explosive eruption of the Hunga Tonga volcano  on January 15, 2022, disturbances were observed at a distance of about 12000\,km in Northern Tien Shan among the variations of the atmosphere pressure, of telluric current, and of the Doppler frequency shift of ionospheric signal. At 16:00:55\,UTC a pulse of atmospheric pressure was detected there with a peak amplitude of 1.3\,hPa and propagation speed of 0.3056\,km/s, equal to the velocity of Lamb wave. In the variations of the Doppler frequency shift, the disturbances of two types were registered on the 3212\,km and 2969\,km long inclined radio-paths, one of which arose as a response to the passage of a Lamb wave (0.3059\,km/s) through the reflection point of radio wave, and another as reaction to the acoustic-gravity wave (0.2602\,km/s). Two successive perturbations were also detected in the records of telluric current at the arrival times of the Lamb and acoustic-gravity waves into the registration point. According to the parameters of the Lamb wave, an energy transfer into the atmosphere at the explosion of the Hunga Tonga volcano was roughly estimated as 2000\,Mt of TNT equivalent.
\end{abstract}


\end{frontmatter}



\section{Introduction}
An explosive eruption of the Hunga Tonga--Hunga Ha\'{}apai volcano in South-West Pacific, which happened on January~15, 2022, was the first event of current century with the volcanic explosivity index~5.
Unique for this event are the powerful disturbances which were observed in the entire thickness of the atmosphere. Everywhere over the  globe it was detected an appearance of intensive infrasonic, acoustic-gravity, and Lamb waves. On-ground barometers in different parts of the world registered the atmosphere signatures of the  volcano explosion \cite{hungotongokulichkov2022,hungotongomatoza2022,hungotongochen2022}, and the atmospheric Lamb wave bypassed the Earth several times, propagating with the mean velocity of $\sim$0.3\,km$\cdot$s$^{-1}$\cite{hungotongokubota2022,hungotongowright2022,hungotongoheki2022}. Disturbances in the ionosphere caused by the Hunga Tonga explosive eruption were studied by the means of the Global Navigation Satellite System (GNSS) receivers net which is commonly used for  measuring the total electron concentration in the ionosphere \cite{hungotongochen2022,hungotongoheki2022,hungotongothemens2022,hungotongozhang2022}. Together with the ground-based GNSS measurements, the prominent ionospheric effects induced by the Hunga Tonga volcano explosion  were also observed from the satellites of the ICE and GOLD missions situated, correspondingly, both at the low-Earth and geostationary orbits \cite{hungotongoaa}.

Generally, it is known that large volcano eruptions cause ionospheric disturbances of various kinds
\cite{hungotongocahyadi2022,hungotongohekifujimoto2022} which are believed to arise from the upward leakage of the energy of Lamb waves. The energy can be transmitted into the ionosphere through an atmospheric resonance at the frequency of acoustic-gravity oscillations, which stipulates large amplitude of the waves at high altitude \cite{hungotongozhang2022}.
In the case of Hunga-Tonga event the GNSS receivers identified two types of the traveling ionospheric disturbances (TID) which propagated from the epicenter of the explosion: there were two large-scale, and several medium-scale TIDs. The most dominant medium-scale TID was moving with the velocity of about (200--400)\,m$\cdot$s$^{-1}$ and coincided with the disturbance of the near-surface atmosphere pressure \cite{hungotongothemens2022}. In a multi-sensor study of the propagation of ionospheric disturbances from the Hunga-Tonga volcano explosion, which was based on more than 5000~GNSS receivers distributed over the whole globe, it was demonstrated that the ionosphere is a sensitive detector of atmospheric waves and geophysical perturbations \cite{hungotongozhang2022}.

During decades continuous multi-parametric observations of geophysical environment were going on in Northern Tien Shan, at the Radio polygon ''Orbita'' of the Institute of Ionosphere, and at the Tien Shan mountain scientific station of P.\,N.\,Lebedev Physical Institute \cite{thunderour2021iskra,seismoour2022ksf}. The experimental base is situated in a mountainous, seismically active locality and encompasses a complex of measuring equipment for simultaneous investigation of the processes which take place in the lithosphere, atmosphere, and ionosphere. The eruption of the Hunga Tonga volcano was one of the most powerful explosive volcanic events of modern era, the disturbances of which were distributed from the lithosphere to the heights of the ionosphere and spread even into the near space. This event occurred a unique natural experiment of such a strong impact on the environment, which was a motif to search for its geophysical consequences at the detectors of the Tien Shan scientific complex.

In the present study the response of the ionosphere to the Hunga-Tonga explosive eruption event was investigated at the distances of $(11-12)\cdot10^3$\,km from the volcano using the method of continuous variation monitoring of the Doppler frequency shift of ionospheric signal on inclined radio path \cite{dopplernazyf1,dopplernazyf2}. The method of Doppler sounding, primarily applied in this work, has demonstrated its efficiency and high sensitivity in a number of studies of ionospheric disturbances which accompany the powerful explosions,  earthquakes, solar flares, and geomagnetic storms \cite{drobzhevkrasnov1978,drobzhevzhelesnyak1987,introref29,introref27,seismoour2022coupling,hungotongosalikhov2020}. Together with the Doppler signal, the most convincing effects of the Hunga Tonga volcanic explosion were found among the measurements of the atmosphere pressure and telluric current, which data are also discussed in the article.

\section{Experimental technique}
Investigation of the response of geophysical fields on the Hunga Tonga--Hunga Ha\'{}apai volcano eruption was carried out at the two high altitude experimental sites: the Radio polygon ``Orbita'' of the Institute of Ionosphere (N43.05831,\,E76.97361; 2750\,m above the sea level), and the Tien Shan mountain scientific station of the Lebedev Physical Institute (N43.03519,\,E76.94139; 3340\,m a.s.l.). Both sites are located  at the territory of the Republic of Kazakhstan in Tien Shan mountains, 2.9\,km apart from each other, and at a distance of 12948\,km from the Hunga Tonga volcano island.

\paragraph{Barometric pressure}
Continuous monitoring of the atmosphere pressure proceeds at the Tien Shan mountain station using a MSB181 type digital barometer (LLC ``MicroStep-MIS'', Russia) which permits to measure the pressure in the range of (600--1100)\,hPa with precision of $\pm$0.3\,hPa. The barometric data with one-minute time resolution are accessible in real time  at an Internet site of the Tien Shan mountain station \cite{tieneng}.

\paragraph{The Doppler frequency shift of ionospheric signal on inclined radio path}
For detection of the ionosphere response to the volcano eruption it was used a hard- and software complex of Doppler measurements which is based on the phase-locked loop (PLL) operation principle and permits to measure the Doppler shift of a larger amplitude radio-signal under conditions of a multipath signal propagation \cite{dopplernazyf1}. Within a 15\,Hz wide holdoff band of the PLL loop the non-linearity of the frequency conversion characteristic equals to 0.46\

\begin{figure*}
{\centering
\includegraphics[width=0.55\textwidth, trim=20mm 0mm 20mm 0mm]{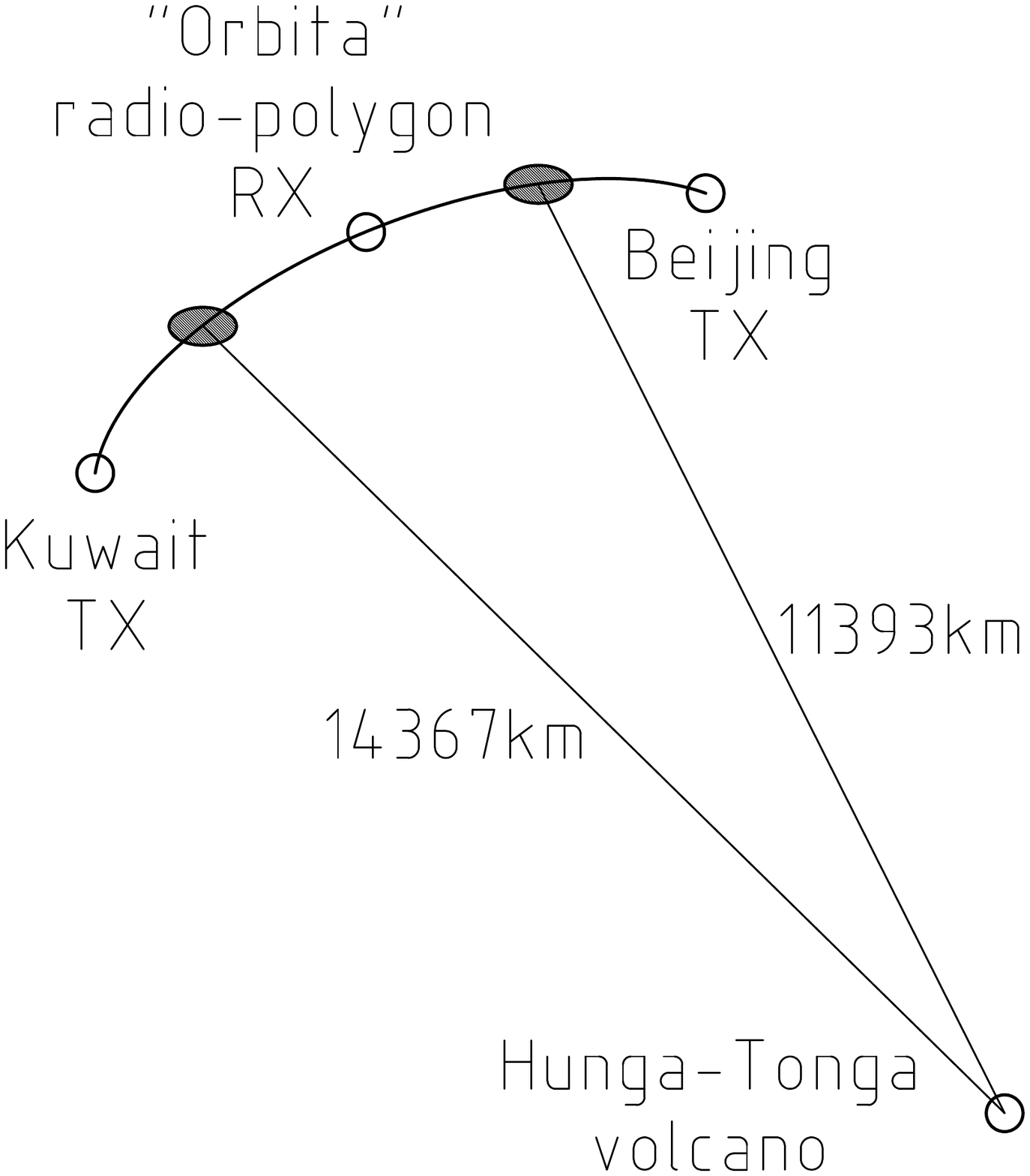}
\includegraphics[width=0.44\textwidth, trim=50mm 0mm 50mm 0mm]{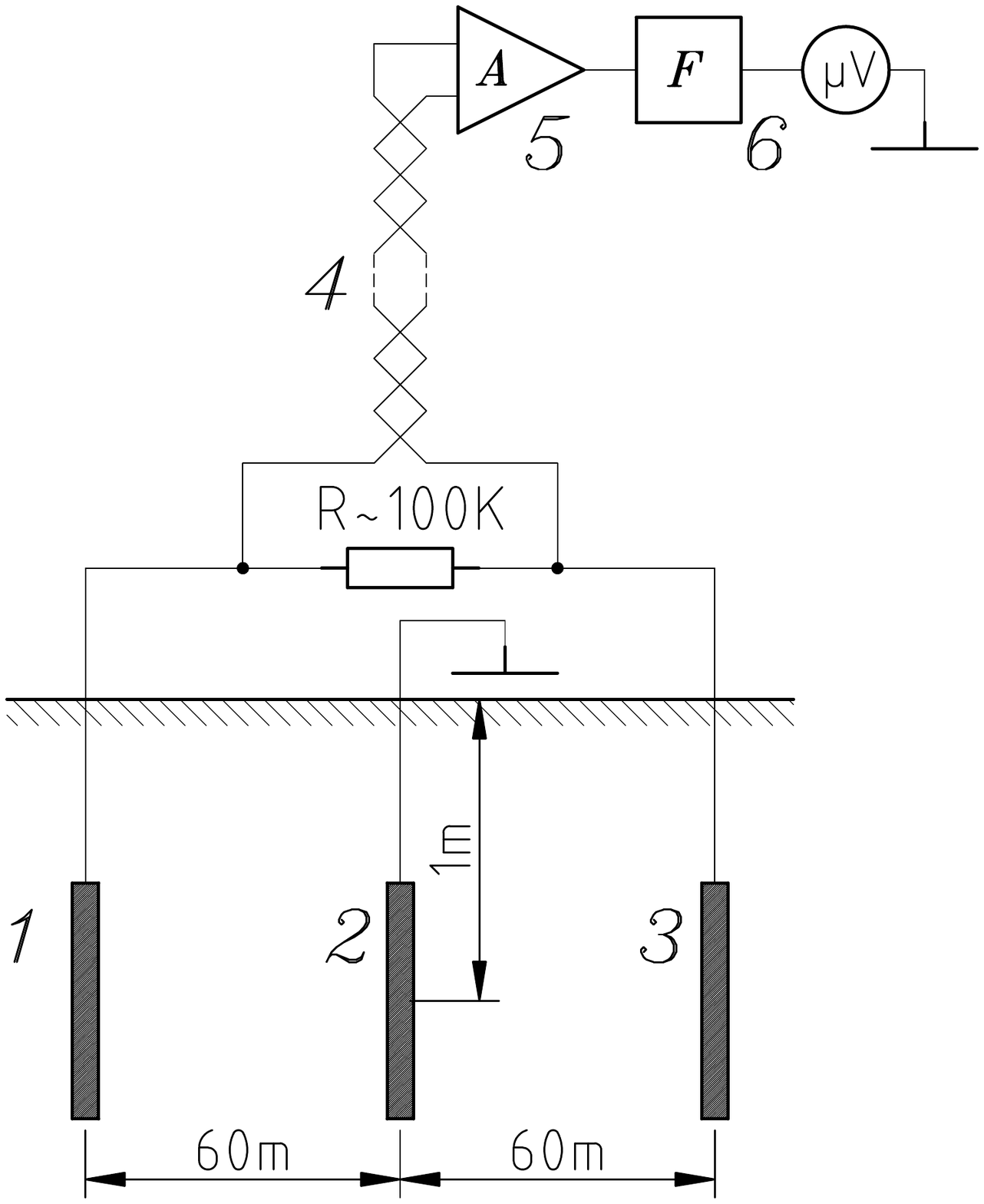}
\caption{Left: the disposition scheme of the radio paths used in the Doppler sounding experiment relative to the Hunga Tonga volcano; \textit{(TX)}---the radio-transmitters, \textit{(RX)}---the receiver at the Radio Polygon ''Orbita''. Two hatched ellipses indicate the projection of the radio wave reflection points onto the Earth. Right: the scheme of the equipment for the measurement of telluric current at the Tien Shan mountain station: \textit{(1),(2),(3)}---three buried electrodes, \textit{(4)}---a $\sim$20\,m long pair of twisted wires, \textit{(5)}---a differential amplifier, \textit{(6)}---an active filtering scheme.
\label{figifig1}}
}
\end{figure*}

In the present experiment, monitoring of the Doppler frequency shift was made over the two inclined radio paths, Beijing---Radio polygon ``Orbita'' (the length $d=3200$\,km and basic frequency $f=7275$\,kHz), and Kuwait---Radio polygon ``Orbita'' ($d=3950$\,km, $f=5860$\,kHz).
Left graph in Figure\,\ref{figifig1} illustrates the mutual disposition of the both mentioned radio paths and of the Hunga Tonga volcano. The distance from the volcano to the projection on the Earth of the point of radio wave reflection equals to 11393\,km for the radio path Beijing---Radio polygon ``Orbita'', and to 14367\,km for the Kuwait---Radio polygon ``Orbita'' path. Both distances were estimated using the Garmin MapSource program.

\paragraph{Measurements of telluric current}

The point of the telluric current measurement is situated at the Tien Shan mountain station, in a locality far away from any industrial sources of electromagnetic interference \cite{our2021_habarlar_telluric}. The measuring equipment consists of three lead electrodes of rectangular shape, 410$\times$80$\times$10\,mm$^3$, buried in vertical position at a depth of 1\,m under the surface of the ground, such that the distance between the two outer electrodes equals to 120\,m. As it is shown in the right plot of Figure\,\ref{figifig1}, the signal of the current is gathered from the two side electrodes, while the middle one is used as a zero point of the electric measurement scheme. For suppression of in-phase interferences the signal from the electrodes is connected to the measurement equipment through a twisted pair of wires and a differential amplifier. Registration of the weak electric oscillations induced on the electrodes by telluric current proceeds in the range of extremely low frequencies, (0--20)\,Hz.

For selection of the weak electric signals with typical amplitude of a few tens---hundred of microvolts it is necessary to eliminate the stray of the industrial power main from the useful signal of telluric current. In the considered experiment, a notch filter was applied for a $\sim$40\,dB rejection  of the industrial 50\,Hz interference. Main filtration of the input signal was made by a 12-order Chebyshev low-pass filter with a cutoff frequency at 32\,Hz and a $\geqslant$80\,dB suppression of high-frequency oscillations above the cutoff. The active filtering scheme was built on the basis of precision low-noise operational amplifiers OPA-27GP of Texas Instruments production. The final registration of the purified signal was made by a 12-bit ADC continuously operating with the digitization speed of 80\,sps \cite{our2021_habarlar_telluric}.

\section{Results and discussion}

\subsection{Anomaly of atmospheric pressure}
\begin{figure*}
{\centering
\includegraphics[width=\textwidth, trim=0mm 0mm 0mm 0mm]{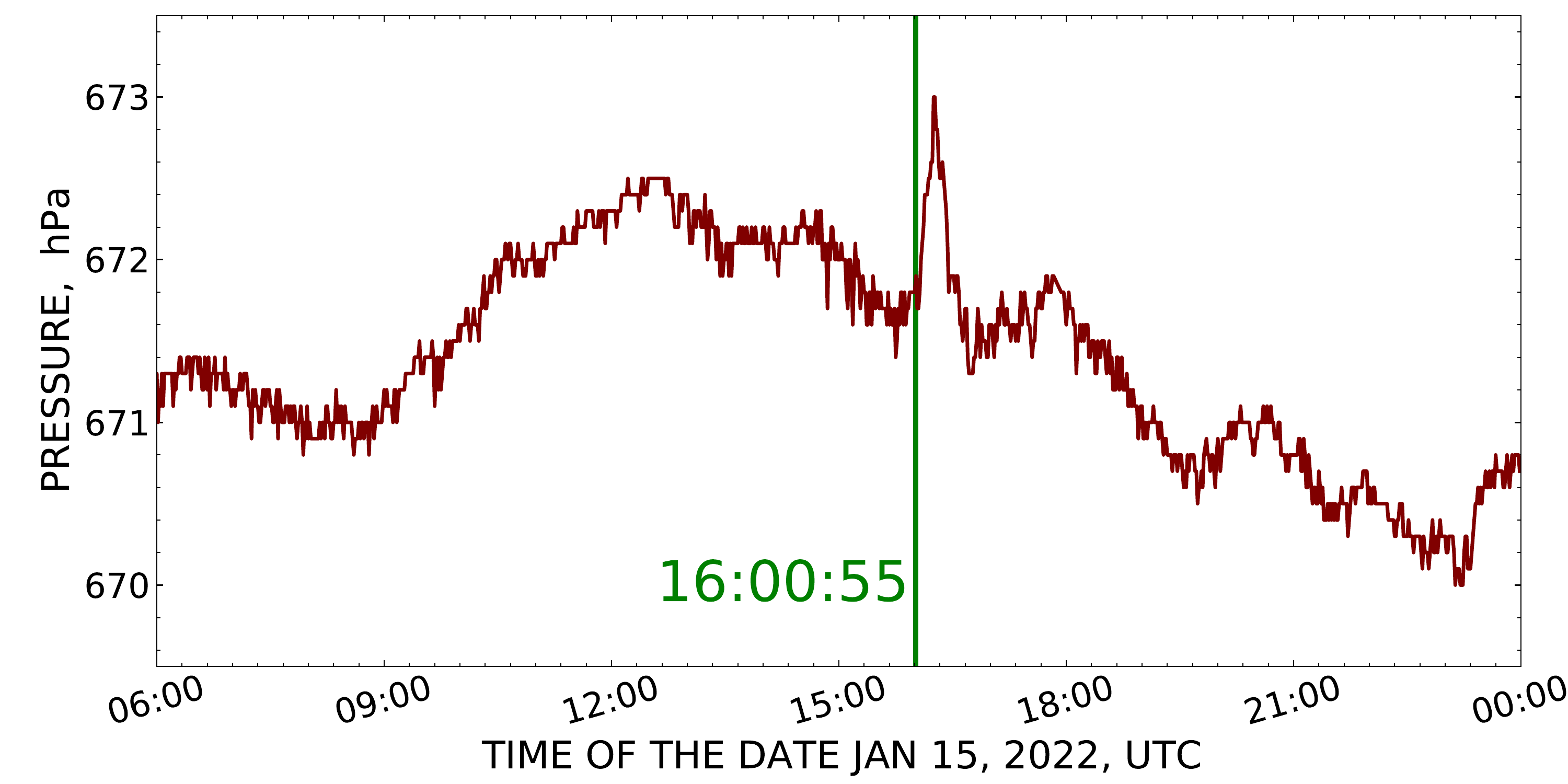}
\caption{The pulse of the atmosphere pressure registered at the Tien Shan mountain station on January 15, 2022.}
\label{figipressi}}
\end{figure*}

On January 15, 2022, in the day of the Hunga Tonga volcano explosive eruption, the barometer located at the territory of the Tien Shan mountain station registered a short time anomalous pulse of the atmosphere pressure, as illustrated by Figure\,\ref{figipressi}. As it follows from this plot, the peak amplitude of pressure growth above the preceding level of undisturbed slow variation was about (1.3$-$1.4)\,hPa, and the whole effect lasted about (25$-$30)\,min. The start of the atmosphere pressure pulse was detected at the moment of 16:00:55\,UTC, nearly 12\,h after the Hunga Tonga volcano  explosion.

\begin{figure*}
{\centering
\includegraphics[width=\textwidth, trim=0mm 5mm 0mm 0mm]{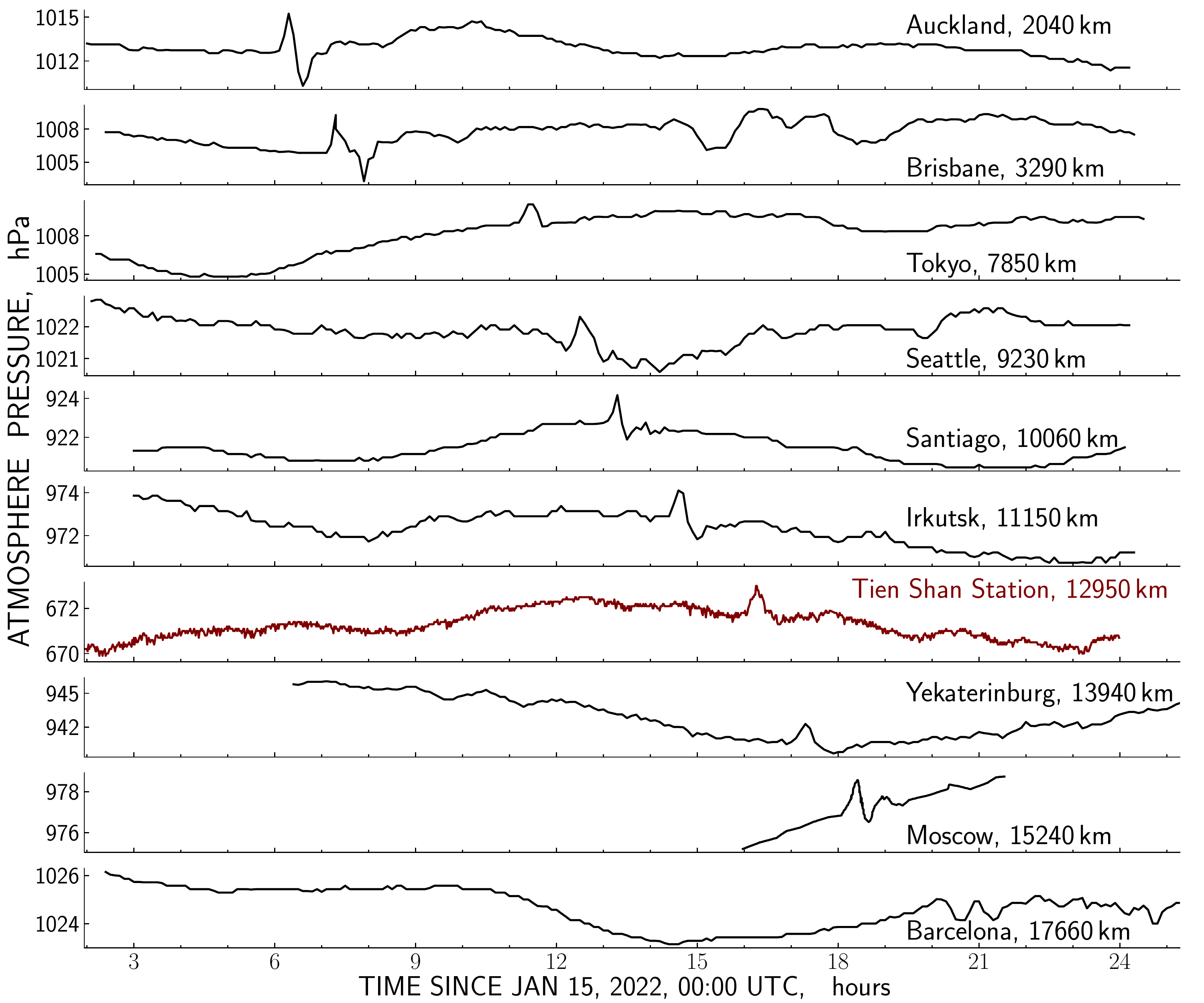}
\\~\\
\includegraphics[width=0.55\textwidth, trim=0mm 5mm 0mm 0mm]{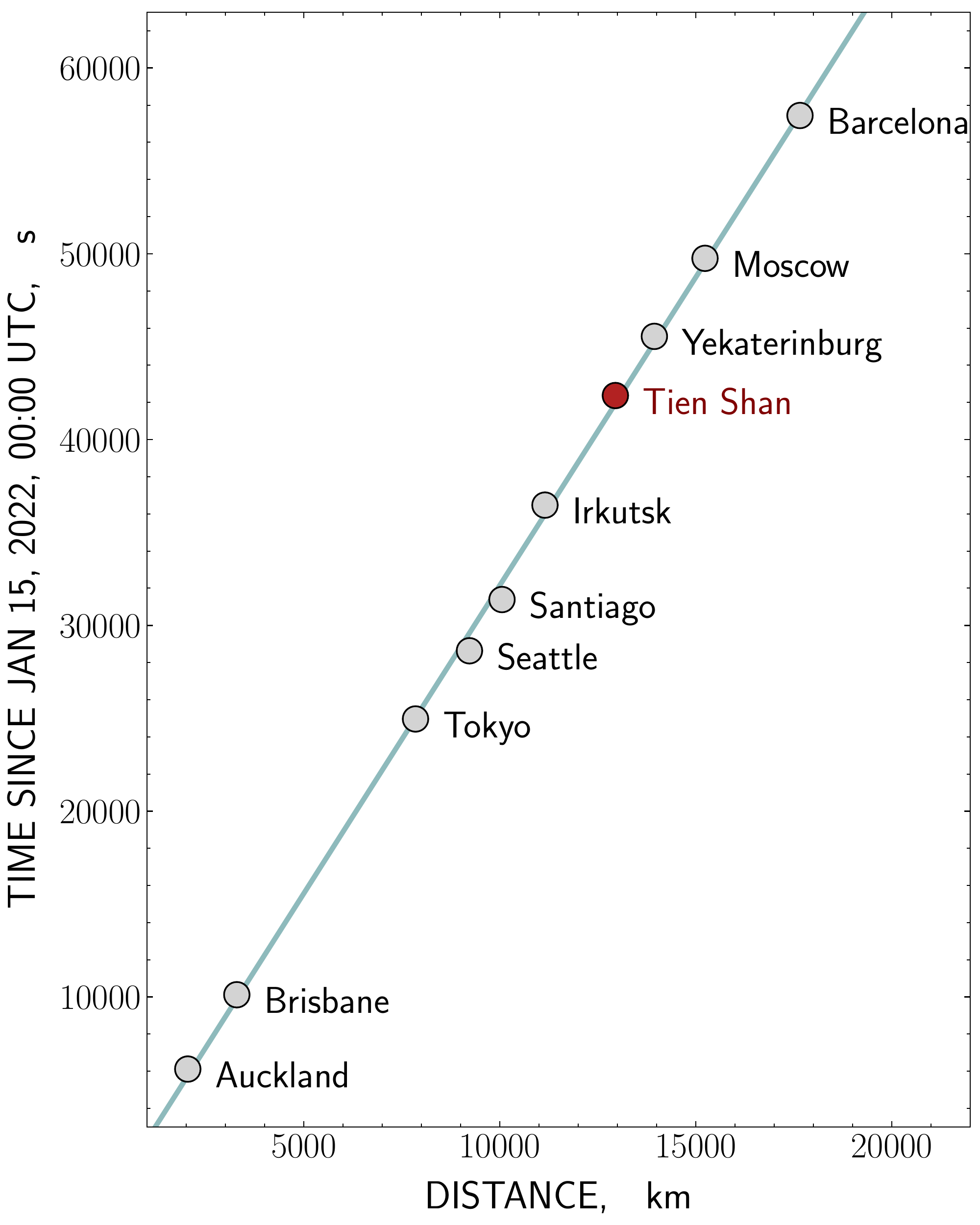}
\caption{Top: anomalous pulses of the atmosphere pressure detected at various distances from the Hunga Tonga volcano on January~15, 2022. Bottom: the dependence of the observation time of the atmosphere pressure pulse on the distance from the volcano.
}
\label{figibaro}}
\end{figure*}

\begin{table}
\begin{center}
\caption{Propagation of the pulse of atmosphere pressure as detected on January~15, 2022 at various distances from the Hunga Tongo volcano.}
\label{tabipressi}
\setcellgapes{1pt} 
\makegapedcells    

\begin{tabular*}{\columnwidth}{@{\extracolsep{\fill}}c|c|c|c|c|c}
\hline
\hline
\makecell{observation\\point} &
\makecell{geographical\\coordinates} &
\makecell{distance,\\km} &
\makecell{arrival\\of pressure\\pulse,\\UTC} &
\makecell{propagation\\time,\\s} &
\makecell{propagation\\speed,\\km$\cdot$s$^{-1}$}
\\
\hline
\hline

\makecell{Auckland}&
\makecell{S36.83970\\E174.82843}&
\makecell{2040}&
\makecell{05:56:44}&
\makecell{6119}&
\makecell{0.3334}
\\
\hline

\makecell{Brisbane}&
\makecell{S27.46778\\E153.02806}&
\makecell{3291}&
\makecell{07:03:24}&
\makecell{10119}&
\makecell{0.3252}
\\
\hline

\makecell{Tokyo}&
\makecell{N36.23000\\E140.18000}&
\makecell{7850}&
\makecell{11:10:47}&
\makecell{24962}&
\makecell{0.3145}
\\
\hline

\makecell{Seattle}&
\makecell{N47.64236\\W122.33348}&
\makecell{9226}&
\makecell{12:11:57}&
\makecell{28632}&
\makecell{0.3222}
\\
\hline

\makecell{Santiago}&
\makecell{S33.31831\\W70.68514}&
\makecell{10056}&
\makecell{12:57:55}&
\makecell{31390}&
\makecell{0.3241}
\\
\hline

\makecell{Irkutsk}&
\makecell{N52.27000\\E104.45000}&
\makecell{11152}&
\makecell{14:22:37}&
\makecell{36472}&
\makecell{0.3058}
\\
\hline

\makecell{\textbf{Tien Shan Station}}&
\makecell{\textbf{N43.04361}\\ \textbf{E76.94139}}&
\makecell{\textbf{12948}}&
\makecell{\textbf{16:00:55}}&
\makecell{\textbf{42370}}&
\makecell{\textbf{0.3056}}
\\
\hline

\makecell{Yekaterinburg}&
\makecell{N56.85400\\E60.64400}&
\makecell{13943}&
\makecell{16:54:00}&
\makecell{45555}&
\makecell{0.3061}
\\
\hline

\makecell{Moscow}&
\makecell{N55.76660\\E37.57324}&
\makecell{15235}&
\makecell{18:04:09}&
\makecell{49755}&
\makecell{0.3061}
\\
\hline

\makecell{Barcelona}&
\makecell{N41.65060\\E2.44564}&
\makecell{17658}&
\makecell{20:12:08}&
\makecell{57443}&
\makecell{0.3074}
\\
\hline

\end{tabular*}

\end{center}
\end{table}

\begin{table}
\begin{center}
\caption{Parameters of the ionospheric disturbance propagation on the radio paths Beijing---Radio polygon ``Orbita'' and Kuwait---Radio polygon ``Orbita''.}
\label{tabidoppli}
\setcellgapes{1pt} 
\makegapedcells    

\begin{tabular*}{\columnwidth}{@{\extracolsep{\fill}}c|c|c|c|c|c}
\hline
\hline
\makecell{radio path} &
\makecell{geographical\\coordinates\\of the\\radio wave\\reflection\\point} &
\makecell{distance\\to the\\reflection\\point,\\km} &
\makecell{time moment\\of the\\ionospheric\\anomaly,\\UTC} &
\makecell{propagation\\time,\\s} &
\makecell{propagation\\speed,\\km$\cdot$s$^{-1}$}
\\
\hline
\hline

\makecell{Beijing---\\``Orbita''}&
\makecell{N43.66260\\E96.75869}&
\makecell{11393}&
\makecell{14:35:21}&
\makecell{37236}&
\makecell{0.3059}
\\
\hline

\makecell{Kuwait---\\``Orbita''}&
\makecell{N36.89300\\E60.39300}&
\makecell{14367}&
\makecell{19:35:21}&
\makecell{55241}&
\makecell{0.2602}
\\
\hline
\end{tabular*}

\end{center}
\end{table}

Similarly, anomalous short-time pulses of the atmosphere pressure were detected in the same day of the Hunga Tonga volcanic event by the barometric equipment installed in the various points of the globe. A number of the atmospheric pressure time series registered at that time, including the record of the Tien Shan station, are presented in the upper panel of Figure\,\ref{figibaro}. Except the Tien Shan station data, the information for these plots was obtained from an open access database \cite{pressureinfo}. It is seen, that everywhere the detected pulses of the atmosphere pressure had similar shape and duration, which varied only slightly between the observation points.

In the bottom panel of Figure\,\ref{figibaro} the moments of the atmosphere pressure pulses are plotted in dependence on the distance between the observation point and the Hunga Tonga volcano. The distances for the latter plot were defined on the basis of the geographical coordinates of corresponding points by the Garmin MapSource program. With account to these distances and the time delays of the pressure pulses it was calculated the average propagation velocity of the atmosphere disturbance, which values are listed in Table\,\ref{tabipressi}.
As it follows from the data presented in Figure\,\ref{figibaro} and Table\,\ref{tabipressi}, the maximum velocity was observed in the nearest registration points, \ie in Auckland and Brisbane, and only starting since the distances above $\sim$$10^4$\,km the atmosphere pressure pulse was propagating in the SE--NW direction with a practically constant speed of 0.306\,km$\cdot$s$^{-1}$.

Positive anomalies of the atmosphere pressure which were moving with the velocity of $\sim$0.3\,km$\cdot$s$^{-1}$ from SE to NW over the territory of Japan and lasted in total about 20\,min were also reported in \cite{hungotongoheki2022}. According to \cite{hungotongokanamori}, generally such pulses of atmospheric pressure are brought by a Lamb wave which propagates with the speed of sound ($\sim$0.3\,km$\cdot$s$^{-1}$) along the surface of the Earth.

It should be noted that Barcelona is situated only at a $\sim$2100\,km distance from the antipode point relative to the Hunga Tonga volcano. This may be the reason why the atmosphere pressure pulse plotted for this station in Figure\,\ref{figibaro} has a maximally complicated shape with two peaks and double duration, which features can be explained by successive registration of the direct and antipodal Lamb waves moving from the volcano explosion point.

The moment of the atmosphere pressure peak observation at the Tien Shan station, 16:00:55\,UTC, agrees well with the general trend of the world data, and corresponds to the propagation speed of a surface Lamb wave from the Hunga Tonga volcano explosion. This agreement permits to identify the observed anomaly of atmospheric pressure as an effect also connected with the Hunga Tonga volcanic event.

\subsection{The Doppler frequency shift of ionospheric signal}

\begin{figure*}
{\centering
\includegraphics[width=0.9\textwidth, trim=0mm 0mm 0mm 0mm]{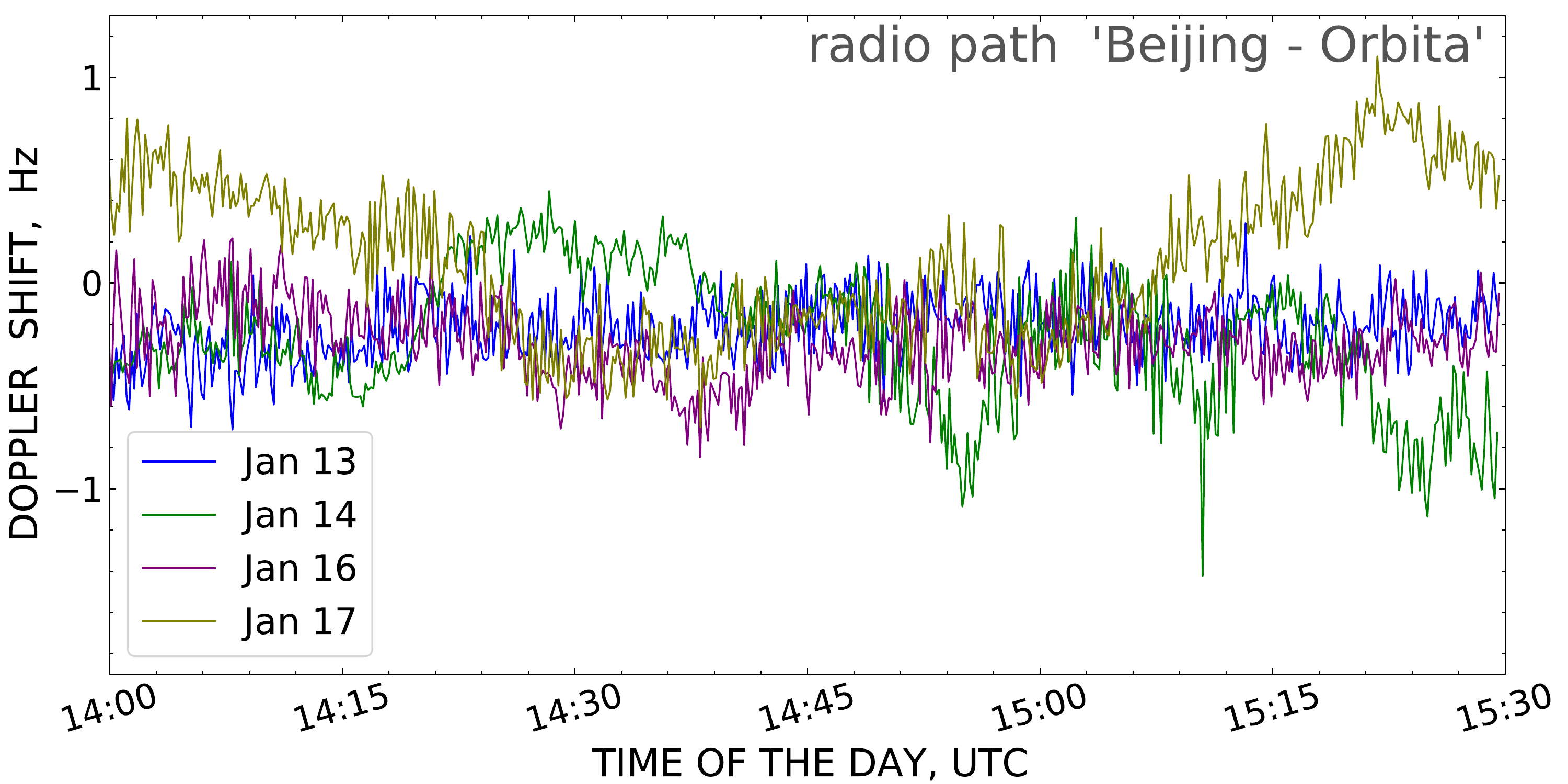}
\includegraphics[width=0.9\textwidth, trim=0mm 0mm 0mm 0mm]{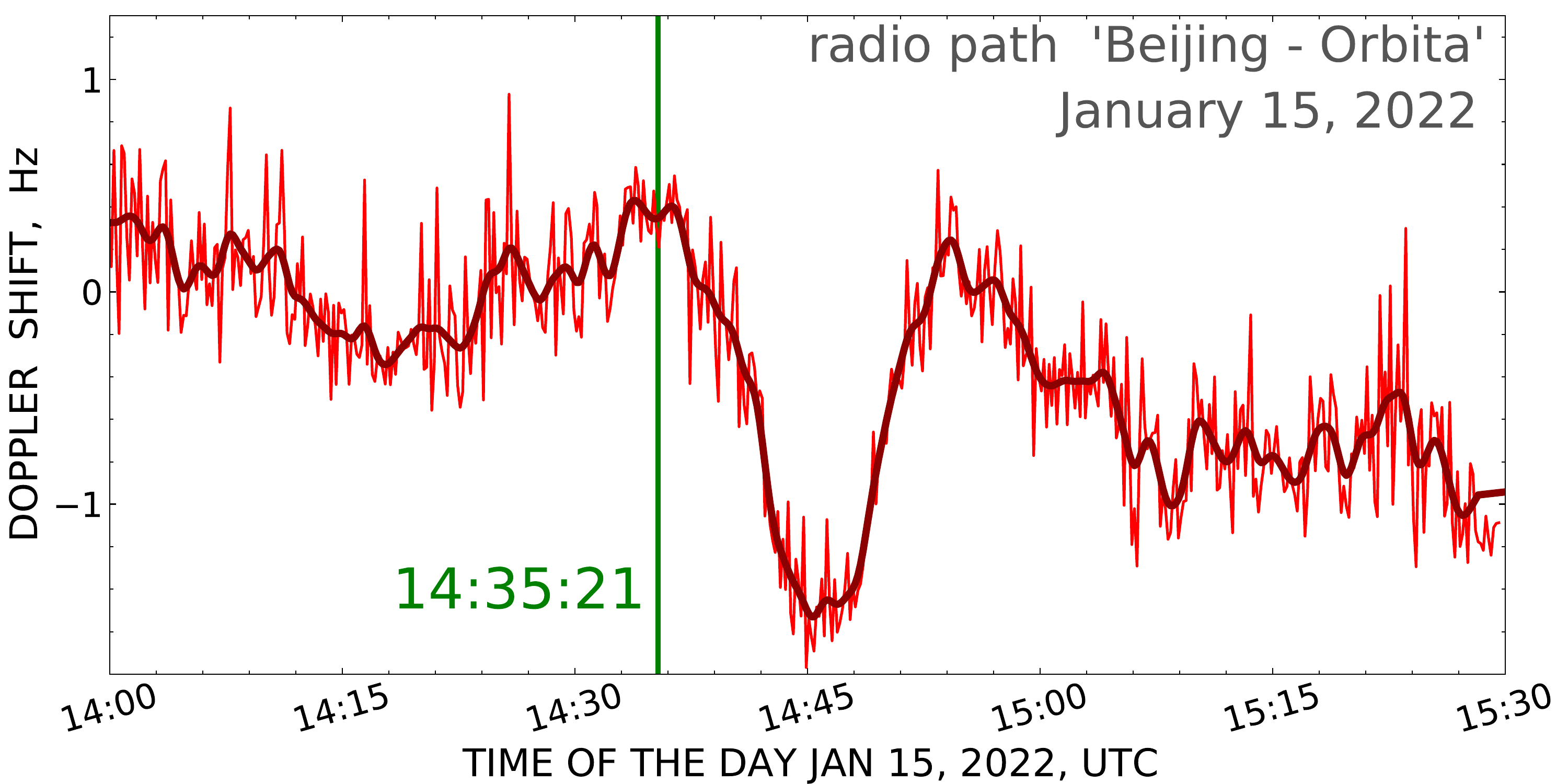}
\includegraphics[width=0.9\textwidth, trim=0mm 0mm 0mm 0mm]{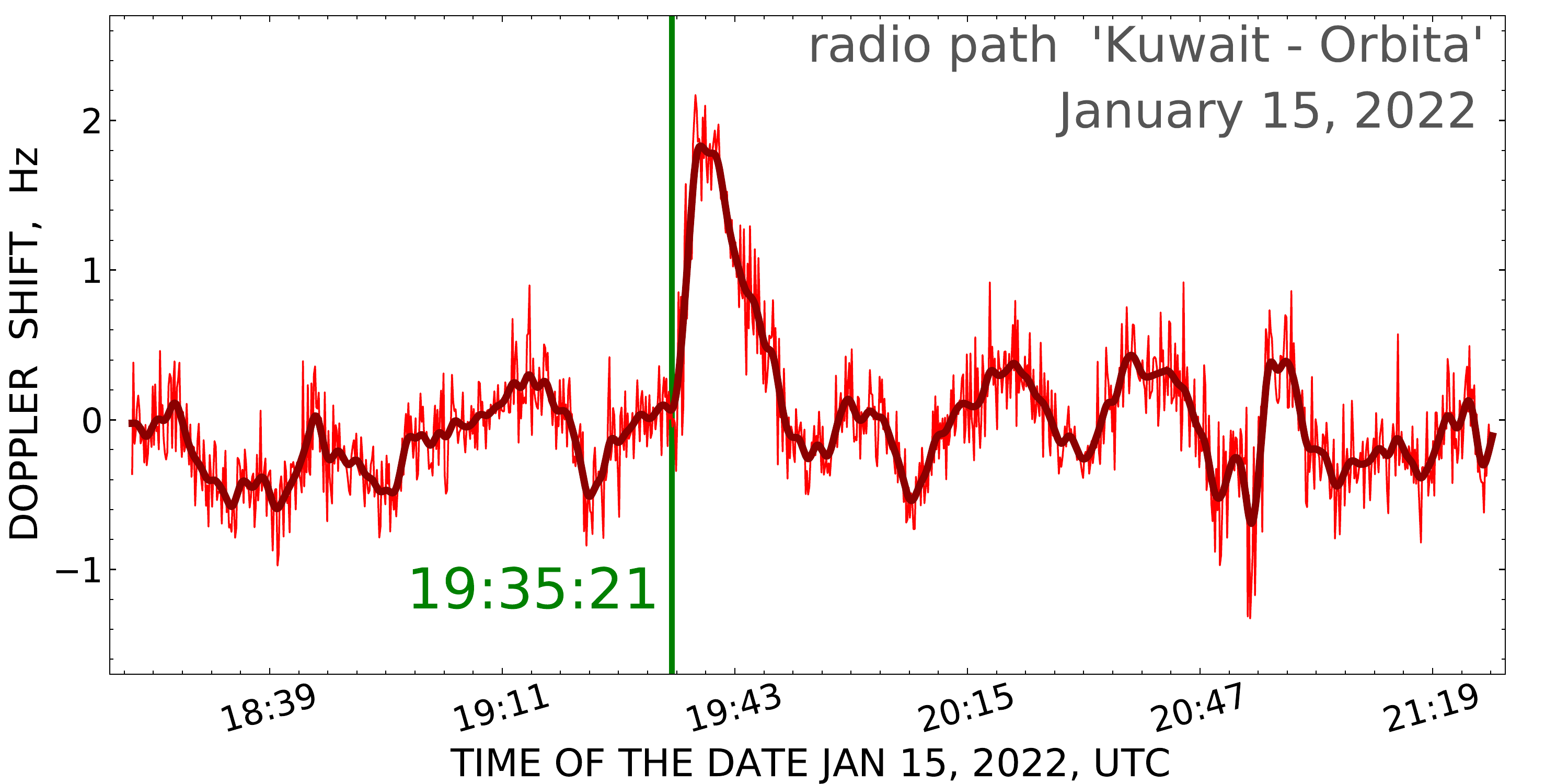}
\caption{
Upper panel: sample time series of the Doppler frequency shift of ionospheric signal registered on the radio path Beijing---Radio polygon ``Orbita'' in the days January 13, 14, 16, and 17, 2022. The middle and bottom panels: the Doppler shift data measured on the paths Beijing---Radio polygon ``Orbita'' and Kuwait---Radio polygon ``Orbita'' on January 15, 2022. Thin lines correspond to the original measurement data of the Doppler frequency, bold lines---to the same data after their smoothing with a 10~points running average filter.
}
\label{figidoppli}}
\end{figure*}

The Doppler frequency shift of ionospheric signal at the Radio polygon ``Orbita'' is registered on the two radio paths, Beijing---Radio polygon ``Orbita'' and Kuwait---Radio polygon ``Orbita''. The optimal radio frequencies for the day- and night time are chosen in accordance with the Short Wave  Radio Frequencies BBC Catalog \cite{radioinfo}. The geometric disposition of both paths relative to location of the Hunga Tonga volcano is illustrated by a chart in Figure\,\ref{figifig1}.

Continuous monitoring of Doppler frequency in the day of the Hunga Tonga volcano explosion, January 15, 2022, made it possible to reveal two disturbances in the ionosphere, which are illustrated by the plots in Figure\,\ref{figidoppli}.

The upper plot of Figure\,\ref{figidoppli} demonstrates a set of superimposed time series of the Doppler frequency shift of ionospheric signal which were obtained at the radio path Beijing---Radio polygon ``Orbita'' on the dates of 13, 14, 16, and 17~January, \ie both immediately on the eve, and a few days after the considered volcanic event. It is seen in this plot, that the undisturbed daily records of Doppler data do not differ considerably between the days, and generally overlap within the limits $\pm$1\,Hz.

In contrast to the regular background oscillations  observed in the preceding days, on January~15, 2022 it was detected a considerable negative disturbance of  Doppler frequency with an amplitude of about 2\,Hz and duration of 1073\,s, which is shown in the middle plot of Figure\,\ref{figidoppli}.
This disturbance appears at the time of 14:35:21~UTC, \ie 37236\,s (10.3\,h) after the explosion moment of the Hunga Tonga volcano.
With account of the distance between the volcano and the reflection point of radio waves at the Beijing---Radio polygon ``Orbita'' radio path, 11393\,km, the propagation velocity of the atmospheric disturbance in the ionosphere can be calculated as 0.3059\,km$\cdot$s$^{-1}$. As it was discussed above, practically same value, 0.3056\,km$\cdot$s$^{-1}$, was obtained at the Tien Shan mountain station for the speed of the pulse of atmosphere pressure which corresponded to the propagation of the Lamb wave from the volcano explosion (see Table\,\ref{tabipressi}).

The observed time correlation is an evidence, that the Lamb wave propagating over the whole thickness of the atmosphere had initiated disturbances in the ionosphere as well, which revealed themselves in the record of the Doppler frequency shift.
As additional argument in favour of this conclusion the ICON-MIGHTI and Swarm satellite observations \cite{hungotongoharding} can be mentioned of the penetration of Lamb waves at the heights of (90--300)\,km into the ionosphere.

The second anomaly in the Doppler frequency shift of ionospheric signal found on January~15, 2022 was registered at the time of 19:35:21\,UTC, \ie 55241\,s (15.3\,h) after the Hunga Tonga volcano explosion, when the Doppler sounding equipment had just switched into operation with the Kuwait---Radio polygon ``Orbita'' radio path. The corresponding time series of Doppler data is presented in the bottom plot of Figure\,\ref{figidoppli}.  It is seen there a disturbance with a maximum frequency shift of 1.84\,Hz and a 1270\,s duration, followed by a sequence of undulating perturbations with much lesser amplitudes and the quasi-periods of 1223\,s, 1146\,s, and 1053\,s. As known, similar periodicity of 1000\,s order can be observed by propagation of acoustic-gravity waves in the ionosphere \cite{drobzhevkrasnov1979}.

In supposition that the reason of the ionospheric effect detected at 19:35:21\,UTC is also connected with the Hunga Tonga volcano explosion, and with account to the 14367\,km distance from the volcano to the radio waves reflection point at the radio path used, a calculation analogous to above gives the distribution speed of the disturbance 0.2602\,km$\cdot$s$^{-1}$. This estimate corresponds well to the velocity of acoustic-gravity waves \cite{hungotongowright2022}.

The characteristics of the two anomalous effects found during the day January 15, 2022 in the records of the Doppler frequency shift of ionospheric signal are summed up in Table\,\ref{tabidoppli}.

Thus, the data obtained at the Radio polygon ``Orbita'' permit to connect the revealed ionosphere disturbances with successive propagation of the Lamb and acoustic-gravity waves generated by the explosion of the Hunga Tonga volcano. At the same time, the detected effects demonstrate the effectiveness of the applied method of Doppler sounding on inclined radio path for revealing ionospheric disturbances of volcanic origin at a distance of more than $10^4$\,km.

\subsection{The measurements of telluric current}
As it follows from the previous section, powerful atmosphere disturbances, distributing after the explosion of the Hunga Tonga volcano as Lamb waves over the entire atmosphere thickness, initiated perturbations in the ionosphere as well. Resulting modulation of the electric currents in the ionosphere causes geomagnetic effects which may be registered at the ground level in variation of the telluric current. This consideration was the reason for searching for anomalies in the monitoring data of telluric current after the Hunga Tonga explosion event.

\begin{figure*}
{\centering
\includegraphics[width=\textwidth, trim=0mm 0mm 0mm 0mm]{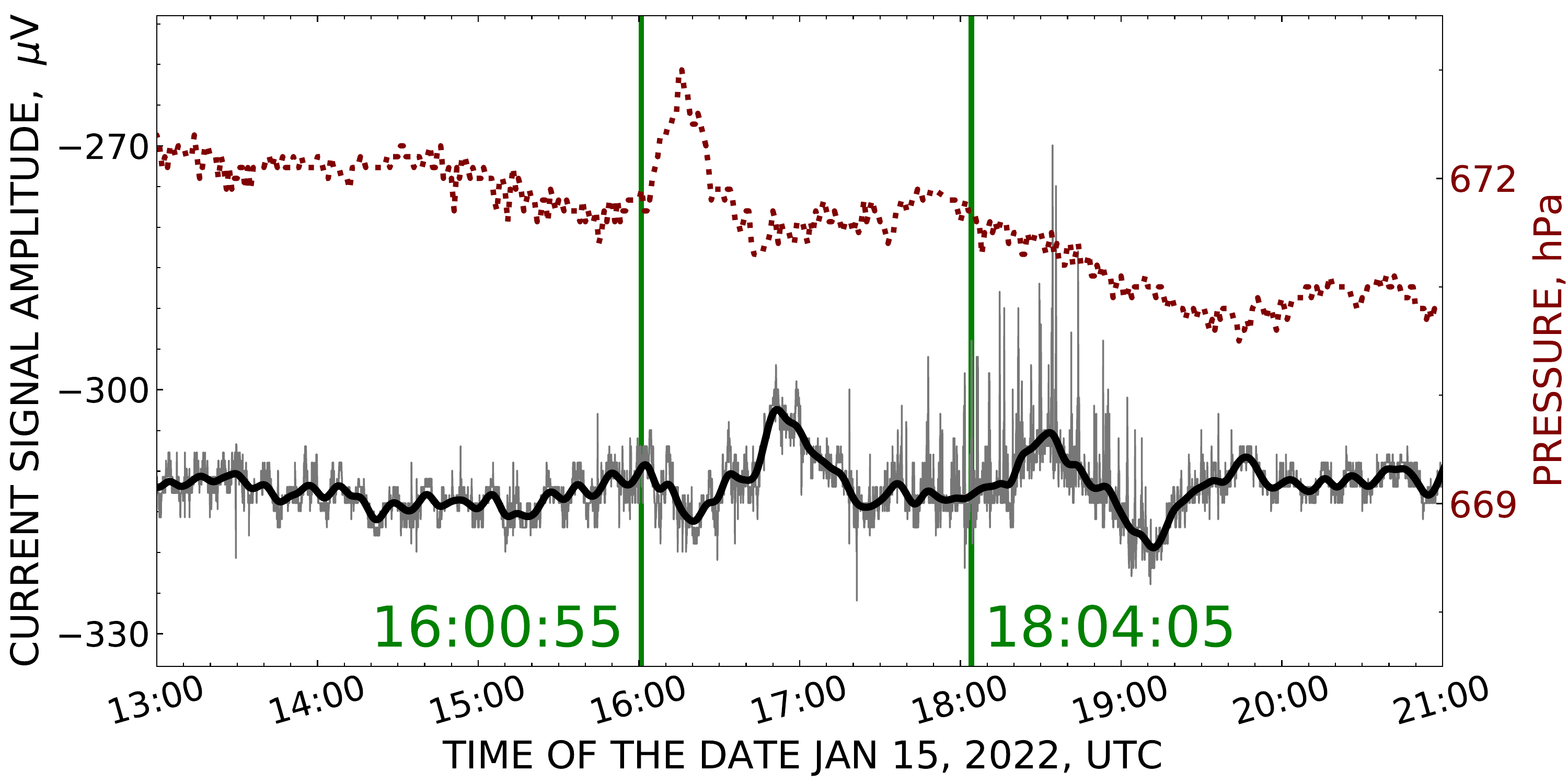}
\caption{Disturbances of the atmosphere pressure and telluric current detected at the Tien Shan station at passage of the acoustic waves from the Hunga Tonga volcano explosion. The dotted line above represents the atmospheric pressure. The thin line in the lower distribution corresponds to the original measurement data of telluric current, while the bold one to the same data smoothed by a running average filter with a 10~points long kernel.}
\label{figitelluri}}
\end{figure*}

In Figure\,\ref{figitelluri} the time series of the atmosphere pressure and of the intensity of telluric current recorded at the Tien Shan station on January~15, 2022 are matched against each other.
As it follows from this figure, at the moment of 16:00:55\,UTC an appreciable (6$-$10)\,$\mu$V decrease appears in the record of telluric current, almost simultaneously with arrival of the atmosphere pressure pulse, which propagated with the 0.3056\,km$\cdot$s$^{-1}$ speed from the Hunga Tonga volcano point.
After the bay-like dip feature in the record it was observed an increase of the telluric current amplitude.

It is worth to note that the decrease of the current lasted about 1800\,s, which is quite comparable with the length of the barometric pressure pulse. The total duration of the whole period of telluric current variations, both of the negative dip and of the subsequent rise, was $\sim$4900\,s.

Another noticeable disturbance in the record of telluric current starts at 18:04:05\,UTC, as it is also seen in Figure\,\ref{figitelluri}. This moment corresponds to the arrival time of the acoustic-gravitation wave from the Hunga Tonga volcano which was moving with the speed of 0.260\,km$\cdot$s$^{-1}$ (see Table\,\ref{tabidoppli}). The second period of the telluric current disturbances lasted about 6800\,s.

The described time correlations may be a reason to suppose an electromagnetic nature of the detected disturbances, at which the modulation of the electric currents in the ionosphere by atmospheric waves may result in the variation of the telluric current measured at the ground level.

\subsection{Evaluation of the energy released into the atmosphere at the Hunga Tonga volcano explosion}

\begin{figure*}
{\centering
\includegraphics[width=\textwidth, trim=0mm 0mm 0mm 0mm]{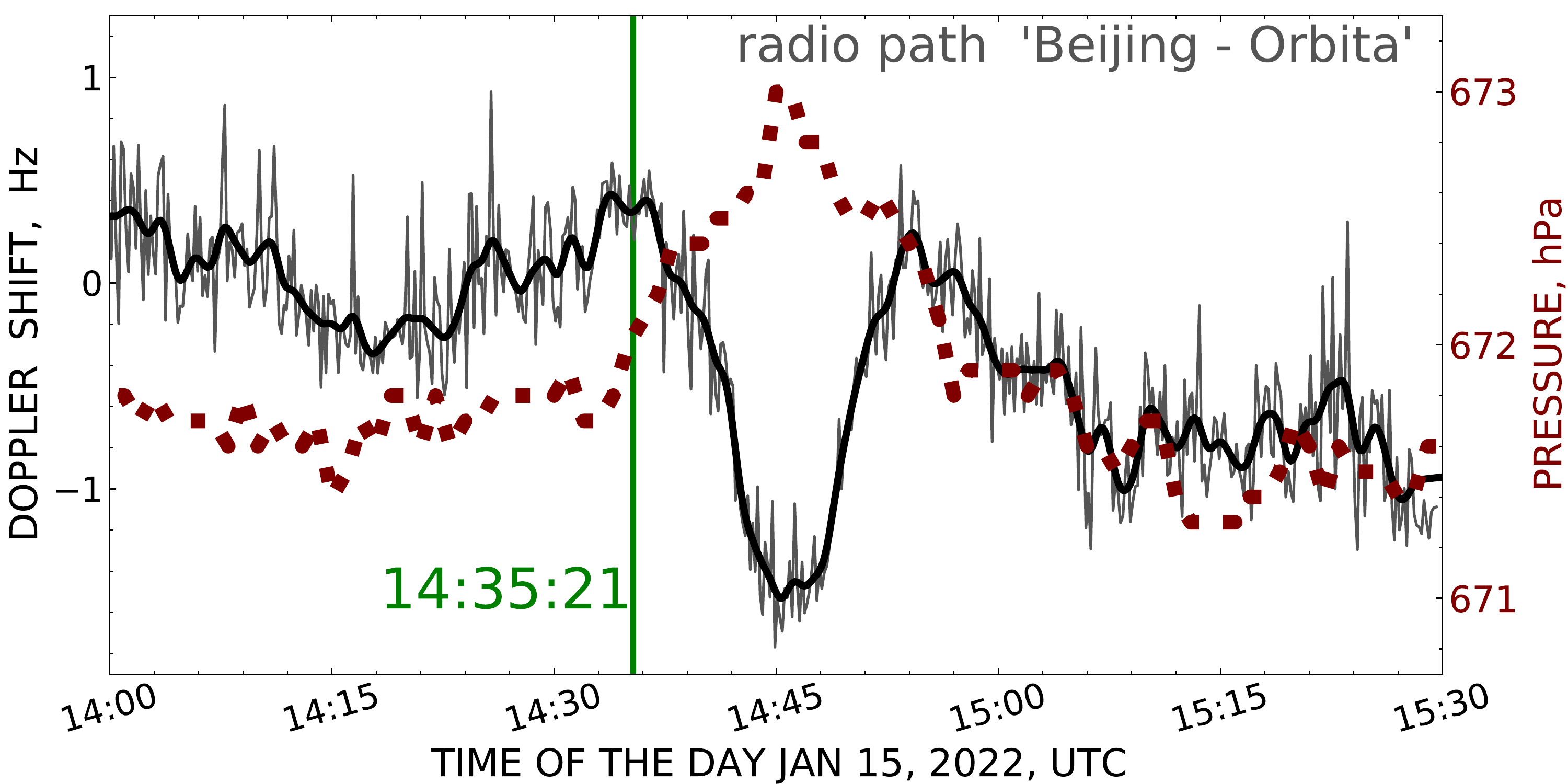}
\caption{Comparison of the data on the Doppler frequency shift at the Beijing---Radio polygon ``Orbita'' radio path registered on January 15, 2022 with the time history of the atmosphere pressure, as it was measured at the Tien Shan station (dotted line). The thin solid line corresponds to the original measurement data of Doppler signal, bold line to the same data after smoothing by a 10~points running average filter. The record of barometric pressure is displaced to $-$1.50\,h along the time axis (see text).}
\label{figidopplioncemore}}
\end{figure*}

In Figure\,\ref{figidopplioncemore} the time history considered above of the Doppler frequency shift of ionospheric signal at the Beijing---Radio polygon ``Orbita'' radio path is superimposed with the record of the atmosphere pressure, as it was written in the date of Hunga Tonga explosion at the Tien Shan mountain station. The barometric data series in this plot is shifted along the horizontal axis to the time of $-$1.50\,h, in correspondence with the propagation speed of the Lamb acoustic wave from the Hunga Tonga explosion, 0.3056\,km$\cdot$s$^{-1}$ (see Table\,\ref{tabipressi}), and the distance difference of 1560\,km between the Hunga Tonga volcano and the Tien Shan station, from the one side, and between the volcano and the reflection point of radio waves on the Beijing---Radio polygon ``Orbita'' radio path, from another.

As it follows from Figure\,\ref{figidopplioncemore}, the duration of the period of anomalous Doppler frequency shift is reasonably close to that of the atmosphere pressure pulse, and the start moments of both effects practically coincide with each other. The distribution speed of the two anomalous effects also seems to be practically identical, about 0.306\,km$\cdot$s$^{-1}$, as it was discussed in the previous paragraphs.
These experimental facts mean that both anomalies originate from the same acoustic disturbance which had arose at a height of about 100\,km and was propagating with the velocity of the surface Lamb wave.
In turn, the specific features of these phenomena permit to estimate the order of energy which was transferred into the atmosphere at the considered explosion event.

Since the average energy of an air molecule is $\frac{3}{2}kT$, the energy density of the atmosphere equals to $\frac{3}{2}nkT$, where $n$ is the concentration of molecules, $T$---temperature, and $k$---the Boltzmann's constant. The barometric pressure $p$ is connected with the molecules concentration by an equation of state $p=nkT$. As a result, the energy density can be expressed through the atmosphere pressure as
\begin{equation}
\varepsilon=\frac{3}{2}p.
\end{equation}
The total energy $\Delta E$ transmitted to the mass of atmosphere by the Hunga Tonga eruption can be found as a product of the disturbed energy density, $\Delta\varepsilon=\frac{3}{2}\Delta p$, to the effective spatial volume of the disturbance region $V_{eff}$:
\begin{equation}
V_{eff}=2\pi r_{\oplus} \sin \left(\frac{r}{r_{\oplus}}\right) H \tau c_s.
\end{equation}
Here, $r_{\oplus}$ is the radius of the Earth; $r$---the distance to the eruption point; $2\pi r_{\oplus} \sin(\frac{r}{r_{\oplus}})$ is the length of the circumference of the disturbance front on the spherical Earth's surface, $\tau$ and $c_s$ are, correspondingly, the time duration of the disturbance and the sound speed; and $H$ is the height of the homogeneous atmosphere, in which the surface Lamb wave concentrates.
Finally, a simple expression for the  transferred energy $\Delta E$ looks as
\begin{equation}
\Delta E=3\pi \Delta p r_{\oplus} \sin\left(\frac{r}{r_{\oplus}}\right) H \tau c_s.
\label{eqeq}
\end{equation}

According to the measurements data presented above, at the Tien Shan mountain station ($r=12948\cdot 10^3$\,m) the average amplitude of the pulse of atmospheric pressure was $\Delta p\approx 0.65$\,hPa, the pressure disturbance lasted $\tau\approx 1000$\,s and had the propagation speed $c_s=306$\,m$\cdot$s$^{-1}$. Putting these values together into Equation\,(\ref{eqeq}), and taking into account that $1$\,hPa~$=100$\,Pa\,[\,J$\cdot$m$^{-3}$\,], $r_{\oplus}=6371\cdot 10^3$\,m, $H=8\cdot 10^3$\,m, and $4.184\cdot 10^{15}$\,J~$=1$\,Mt$_{\text{TNT}}$, the next rough estimate follows for the energy released into the atmosphere by the Hunga Tonga explosive eruption, as expressed in the megatons of TNT equivalent:
\begin{equation}
\Delta E\approx 2\cdot 10^3\text{\,Mt}_\text{TNT}.
\end{equation}

\section{Conclusion}
The measurement equipment installed at the Radio polygon ``Orbita'' and at the Tien Shan mountain scientific station has registered disturbances in the near-surface atmosphere, in the ionosphere, and among the variations of telluric current, which took place after explosion of the Hunga Tonga volcano on January 15, 2022. In this experiment, the monitoring of the atmosphere pressure, of the Doppler frequency shift of ionospheric signal, and of the telluric current in the near-surface layers of the ground was made at a considerable distance, of about $12\cdot10^3$\,km from the volcano. The observations may be summarized as the following.

1. On January 15, 2022, 11\,h, 46\,min, and 10\,s after the volcano explosion, it was detected an anomalous short time pulse of the atmosphere pressure which had an amplitude of 1.3\,hPa and duration of 
about (25$-$30)\,min, and was propagating in the atmosphere with the velocity of Lamb waves, 0.3056\,km$\cdot$s$^{-1}$, as it was registered by the barometer of the Tien Shan mountain station.

2. Continuous monitoring of the Doppler frequency shift of ionospheric signal on the inclined radio paths with the length of 3212\,km (7245\,kHz) and 2969\,km (5860\,kHz) permitted to reveal two different ionosphere disturbances in the day of the Hunga Tonga explosive eruption. The disturbances onset time corresponded to arrival of the atmospheric waves moving with the velocities of 0.3059\,km$\cdot$s$^{-1}$ and 0.2602\,km$\cdot$s$^{-1}$ into the reflection point of radio waves in the ionosphere. Judging by the velocity values, the observed disturbances in the Doppler frequency shift of ionospheric signal arose as a result of successive passage of the Lamb- and acoustic-gravity waves generated by the volcano explosion, and their influence to the ionosphere.

3. In the day of volcano explosion it was found two consecutive disturbances in the variations of telluric current, their appearance being consistent with the passage of the atmospheric waves with velocities of 0.3056\,km$\cdot$s$^{-1}$ and 0.2600\,km$\cdot$s$^{-1}$. Both velocity estimates permit to connect the observed effects with the passage of the Lamb- and acoustic-gravity waves from the volcano explosion across the point of the telluric current registration. Thus, the atmospheric waves, propagating over the whole 
thickness of the atmosphere, cause modulation of the  electric currents in the ionosphere, which induces an electromagnetic response in the telluric current registered at the ground level.

4. The energy transferred into the atmosphere at the explosion of the Hunga Tonga volcano is roughly estimated by the parameters of the Lamb wave as 2000\,Mt of TNT equivalent.





\end{document}